**A model of H-NS mediated compaction of bacterial DNA**

Marc JOYEUX [(*)]

*Laboratoire Interdisciplinaire de Physique (CNRS UMR5588),*
*Université Joseph Fourier Grenoble 1, BP 87, 38402 St Martin d'Hères, France*

and Jocelyne VREEDE [(+)]

*Van't Hoff Institute for Molecular Sciences, University of Amsterdam,*
*P.O. Box 94157, 1090 GD Amsterdam, The Netherlands*

**Abstract** : The Histone-like Nucleoid Structuring protein (H-NS) is a nucleoid-associated protein, which is involved in both gene regulation and DNA compaction. H-NS can bind to DNA in two different ways: in *trans*, by binding to two separate DNA duplexes, or in *cis*, by binding to different sites on the same duplex. Based on scanning force microscopy imaging and optical trap-driven unzipping assays, it has recently been suggested that DNA compaction may result from the antagonistic effects of H-NS binding to DNA in *trans* and *cis* configurations. In order to get more insight into the compaction mechanism, we constructed a coarse-grained model description of the compaction of bacterial DNA by H-NS. These simulations highlight the fact that DNA compaction indeed results from the subtle equilibrium between several competing factors, which include the deformation dynamics of the plasmid and the several binding modes of protein dimers to DNA, *i.e.* dangling configurations, *cis*- and *trans*-binding. In particular, the degree of compaction is extremely sensitive to the difference in binding energies of the *cis* and *trans* configurations. Our simulations also point out that the conformations of the DNA-protein complexes are significantly different in bulk and in planar conditions, suggesting that conformations observed on mica surfaces may differ significantly from those that prevail in living cells.

(*) email : Marc.Joyeux@ujf-grenoble.fr
(+) email : J.Vreede@uva.nl

## INTRODUCTION

The chromosomal DNA of bacteria is folded into a compact body, called the nucleoid, which is distinctly different from the rest of the cytoplasm. Bacteria employ a number of Nucleoid-Associated Proteins (NAPs) that influence the organization of the nucleoid by bending, wrapping, or bridging DNA (1-4). In addition, most of these proteins act as global regulators of gene expression (5,6). The Histone-like Nucleoid Structuring protein (H-NS) is a key player in genome organization by forming bridges between DNA duplexes (7,8). Moreover, H-NS is an important global regulator (regulating several hundreds of genes) and may function as an environmental sensor (9,10) that perceives changes in the growth conditions of the bacterium and facilitates physiological changes required for adaptation to the new conditions. H-NS is a small protein (137 residues, 15.5 kDa), which is functional as a dimer. Each monomer is composed of an N-terminal dimerization domain (residues 1-64) (11,12) connected to a C-terminal DNA-binding domain (residues 91-137) (13) through a flexible linker (14). The dimer can bind to DNA in two different ways, namely in *trans*, by binding to two separate DNA duplexes, and in *cis*, by binding to different sites on the same duplex (7,8,15,16). The current view reflects that H-NS initially binds to specific nucleotide sequences (13,17), followed by additional H-NS dimers binding to adjacent sites on DNA (14,18).

Despite the enormous effort of many research groups, there are several aspects of H-NS that remain controversial or completely unknown (19). Up to now, a complete, high-resolution structure of full-length H-NS, either free or complexed with DNA, has proven elusive. Structural models for the N-terminal dimerization domain have been determined by NMR spectroscopy (11,12) and X-ray crystallography (18), showing that dimerization occurs through the formation of a coiled-coil interaction motif. However, the degree of oligomerization of H-NS in solution is unclear, since dimers, tetramers and larger oligomers have been observed under different conditions (11,20-23). An NMR spectroscopy study on the DNA-binding domain revealed that a specific sequence in the domain binds to the minor groove (13). A genome-wide analysis identified a strong correlation between H-NS binding sites and AT-rich regions on the DNA (24), and in addition, a 10 bp high-affinity H-NS binding site was also discovered in the *proV* promoter (25). Still, the specific mechanisms for recognition of these regions by H-NS remain unclear. Furthermore, the mechanism by which an H-NS dimer determines to bind in *trans* or in *cis* mode is unknown. By probing H-NS/DNA interactions through equilibrium H-NS mediated DNA looping scanning force microscopy imaging and dynamic optical trap-driven unzipping assays, Wiggins *et al.* suggested that observed DNA loops may result from the antagonistic effects of *trans-* and *cis-*binding (15).

Molecular simulations can complement experiments by providing details at high spatial and temporal resolution. For example, coarse-grained Monte Carlo simulations recently indicated that H-NS bridge formation occurs preferentially at planar bends in the DNA (26). In this paper, we aim to shed additional light on the H-NS mediated compaction of DNA by using a Hamiltonian coarse-grained model for H-NS binding to DNA. Such models have proven to be very helpful in understanding *facilitated diffusion*, *i.e.* the various mechanisms by which proteins scan DNA sequences while searching for their target site (27-29). In this model, *trans-* and *cis-*binding are not introduced *a priori*, but follow instead from the explicit interaction energy function between DNA and the proteins. The results presented in this paper highlight the fact that DNA compaction is indeed driven by the subtle equilibrium between several competing factors, including the deformation dynamics of the plasmid and the several binding modes of protein dimers to DNA. Moreover, simulations of DNA constrained to a planar system suggest that the conformations of DNA/H-NS complexes deposited on mica surfaces may differ substantially from those that prevail in living cells.

## METHODS

Construction of the coarse grained model was aimed at mimicking the DNA/H-NS incubation conditions used by Dame *et al.* (7). The Supporting Material gives a detailed description of the model. In brief, it consists of a pUC19 plasmid and 224 H-NS dimers (*i.e.* 1 dimer:12 bp, as in (7)) at a temperature of $T$=298 K (1 $k_{\mathrm{B}}T \approx$2.48 kJ/mol) enclosed in a sphere of radius $R_0 = 0.434$ μm or a half-sphere of radius $R_0 = 0.547$ μm. As in previous work (27-30), DNA is modelled as a chain of beads, where each bead represents 15 DNA base pairs. This level of coarse-graining is reasonable, because studies have shown that an H-NS DNA-binding domain occupies about 15 bp (31). As a consequence, the 2686 bp pUC19 plasmid is modelled as a cyclic chain of 179 beads. To keep the same level of coarse-graining, H-NS dimers are modelled as chains of three beads. The chosen equilibrium distance between two successive beads ( $L_0 = 7.0$ nm) corresponds to the two N-terminal domains of H-NS dimers being arranged in anti-parallel (12) rather than in parallel (11) geometries. Electric charges are placed at the center of each bead ($-12\,\bar{e}$ on DNA beads, $4\,\bar{e}$ on H-NS terminal beads, and $-8\,\bar{e}$ on H-NS central beads, where $\bar{e}$ denotes the absolute charge of the electron) and the DNA molecule and the H-NS dimers interact through the sum of electrostatic terms (Debye-Hückel potentials) and excluded volume terms. The parameter $\chi$ of the excluded volume terms was chosen such that the computed enthalpy change upon a single H-NS molecule binding to DNA matches the experimental value at $T$=298K (31). For a dimer binding to DNA, the computed enthalpy change is $\Delta H = 11.1\ k_{\mathrm{B}}T$ (see Fig. S1 of the Supporting Material), and the enthalpy change of an H-NS dimer binding two DNA sites in the *trans* configuration is twice this value, 22.2 $k_{\mathrm{B}}T$. Since at equilibrium H-NS dimers are assumed to be linear, the enthalpy change of an H-NS dimer binding two DNA sites in the *cis* mode depends on its bending rigidity $G$. We performed simulations with two different values of $G$, namely $G = 2\,k_{\mathrm{B}}T$ and $G = 4\,k_{\mathrm{B}}T$, which correspond to *cis*-binding enthalpy changes of 19.2 $k_{\mathrm{B}}T$ (see Fig. S2 of the Supporting Material) and 16.8 $k_{\mathrm{B}}T$ (see Fig. S3 of the Supporting Material), respectively.

The dynamics of the system was investigated by numerically integrating the Langevin equations of motion for 10 ms with a time step of 20 ps. Two different sets of simulations were performed. The first set of simulations, hereafter labelled 3D, aimed at mimicking the motion of the system in bulk buffer by allowing all molecules to move freely inside the sphere. The second set of simulations, hereafter labelled 2D, aimed at mimicking DNA molecules deposited on cleaved mica surfaces by restricting the motion of H-NS dimers to a half-sphere, and constraining the plasmid to remain in the neighbourhood of the limiting disk. Four 2D simulations were run for each value of $G$ starting from circular plasmid conformations and eight 3D simulations were run for each value of $G$ starting from thermalized plasmid conformations. Analysis of the simulations consisted in counting every 10 μs the number of bound H-NS dimers. An H-NS dimer was considered bound to DNA if at least one bead was within a distance of 2 nm from a DNA bead (see the Supporting Material and Fig. S4 of the Supporting Material for a discussion on the choice of 2 nm for the distance threshold). Also, the nature of the bound H-NS dimers was tracked. A dimer was considered to be bound in *cis* if the two terminal beads were interacting with DNA beads separated by at most two DNA beads. If instead, the H-NS dimer was interacting with two DNA beads separated by (strictly) more than two DNA beads, the H-NS/DNA interaction was counted as a *trans* bond. Finally, the dimer was considered to be dangling, if only one bead of the dimer was interacting with the DNA. Furthermore, the radius of gyration of the plasmid, *i.e.* the square root of the trace of its gyration tensor, was also computed every 10 μs.

## RESULTS AND DISCUSSION

As described in the Methods section, the model consists of a cyclic DNA molecule with 179 binding sites and 224 H-NS dimers, which move under the action of thermal noise and interaction forces, enclosed in a sphere. When an H-NS dimer comes close to a free DNA site, binding can occur between this DNA bead and a terminal bead in the H-NS dimer, due to the electrostatic attraction between the positive charge placed on the H-NS terminal beads and the negative charge placed on the DNA bead. Subsequently, the other, dangling bead of the H-NS dimer can bind to another site on the DNA. If the second site is within three beads of the first one, the H-NS dimer has bound in *cis* mode. Alternatively, the second binding site can be located far away from the first site from a sequence point of view, resulting in an H-NS dimer bound in *trans*. All bonds between H-NS and DNA are reversible, which means that a twice-bound H-NS dimer can return to the dangling configuration and that a dangling H-NS can detach from DNA or slide along the sequence. As only *trans*-binding dimers contribute to DNA compaction (*cis*-binding and dangling dimers contribute to DNA stiffening (16)), the number of *trans*-binding H-NS dimers is the key quantity to measure DNA compaction. In contrast, *cis*-binding dimers tend to render the DNA inactive, as they prevent binding sites to participate in DNA compaction. The dangling dimers have a more versatile role. Even though they do not contribute to compaction at the current time step, their probability to form a *trans*-binding dimer at later time steps may be high.

The motion of DNA plays a significant role in the global dynamics of H-NS dimers in the presence of DNA. The simulations show that thermal noise, introduced by the Langevin equations of motion, is sufficient to severely deform the cyclic plasmid chain, allowing for two sites, widely separated in sequence, to eventually come close in space. Upon two sequentially distant sites approaching each other, one of three events may happen. 1) If no H-NS dimer is bound, the sites will again move apart after a short while. 2) If both sites contain a *cis*-binding or dangling H-NS dimer, both sites will also move apart again. 3) In the case of one free site and one site occupied by a dangling H-NS dimer approaching each other, *trans*-binding can occur. A single *trans*-binding bridge is not very strong by itself and is likely to be broken by further deformation of the plasmid chain due to thermal noise. Nevertheless, a single *trans*-bound dimer keeps the two sites in close vicinity for a substantially longer time. As a consequence, additional *trans*-binding dimers are more likely to form at sites close to the first *trans*-binding dimer, provided that other pairs of free and dangling dimer sites are available. Therefore, under suitable conditions (7) the first *trans*-binding dimer can act as a seed to propagate compaction, a mechanism sometimes referred to as *cooperativity*.

### The effect of the H-NS dimer *cis*-binding energy on DNA compaction

Since there are many interacting possibilities at each step, the final state of the system will be very sensitive to small changes in the description of the interactions. First we investigated the effect of changing the *cis* interaction energy. This energy depends on the bending rigidity $G$. At higher values of $G$, the H-NS dimer is stiffer and less likely to bend. Therefore, we expect that at higher values of $G$ less H-NS dimers will bind in *cis* mode, because bending of the H-NS molecule is required for *cis*-binding. We performed 3D simulations for $G = 2\,k_{\mathrm{B}}T$ and $G = 4\,k_{\mathrm{B}}T$, which correspond to *cis*-binding enthalpy differences $\Delta H$ of 19.2 $k_{\mathrm{B}}T$ and 16.8 $k_{\mathrm{B}}T$, respectively. Fig. 1 shows the time evolution of the number of H-NS dimers bound to DNA for $G = 2\,k_{\mathrm{B}}T$ (bottom plot) and $G = 4\,k_{\mathrm{B}}T$ (top plot), partitioned in dangling dimers, *cis*- and *trans*-binding dimers. For both values of $G$, the average number of dangling proteins is comparable, tending towards a value close to 50 at long times. In contrast, there is a huge difference in the number of *cis*-binding dimers. Indeed, for $G = 2\,k_{\mathrm{B}}T$ the average number of *cis*-binding dimers tends towards 25 or 30 at long times, while this number remains very small, close to 5, for $G = 4\,k_{\mathrm{B}}T$. Conversely, the number of *trans*-binding dimers at long times is close to 5 and 25 for $G = 2\,k_{\mathrm{B}}T$ and $G = 4\,k_{\mathrm{B}}T$, respectively. This indicates that the moderate difference in the value of the *cis*-binding energy is sufficient to induce large changes in the average number of *cis*- and *trans*-binding dimers. In contrast, the time evolution of the fraction of DNA sites occupied by H-NS dimers, shown in Fig. S5 of the Supporting Material, indicates that the value of $G$ has little effect on the occupation of DNA sites by H-NS. Indeed, the fraction of occupied sites after 10 ms is close to 0.65 for both values of $G$.

The huge difference in the number of *cis*- and *trans*-binding dimers for the two different bending rigidities has profound effects on the dynamics of the condensed plasmid, as illustrated in Fig. 2. This figure shows the occupation of DNA sites by *trans*-binding H-NS dimers, which contribute to compaction, for $G = 2\,k_{\mathrm{B}}T$ (bottom plot) and $G = 4\,k_{\mathrm{B}}T$ (top plot). At $G = 4\,k_{\mathrm{B}}T$, the large number of *trans*-binding dimers facilitates many bridges between DNA sites located everywhere in the sequence. As a result, the plasmid molecule acquires a globular shape, which changes little with time, as demonstrated by the almost constant value of its radius of gyration, plotted in Fig. 3 for a single simulation and in the top plot of Fig. S6 of the Supporting Material for eight different simulations. In contrast, for $G = 2\,k_{\mathrm{B}}T$, the small number of *trans*-binding H-NS dimers allows for only a few, short-lived bridges, which form and break with fluctuations in the plasmid conformation (see Fig. 2). As a consequence, the plasmid molecule displays a very rich dynamics, with compact shapes following more loosely packed conformations and *vice versa*. This behaviour is again clearly reflected in the time evolution of the radius of gyration of the plasmid, shown in Fig. 3 for a single simulation and in the bottom plot of Fig. S6 for eight different simulations. Movies S1 and S2, showing the time evolution of the plasmid conformations for the simulations also shown in Figs. 2 and 3, clearly illustrate the differences in the plasmid dynamics.

## A comparison of H-NS mediated compaction in bulk and on a surface

The simulations discussed above mimicked bulk solution conditions. However, in the scanning force microscopy imaging experiments of Dame *et al.* (7), solutions containing DNA/H-NS complexes were deposited on freshly cleaved mica surfaces followed by gently drying of the mica disks. Deposition on a surface plate may significantly alter the conformations displayed by naked DNA in comparison to the bulk solution situation (32). Magnesium ions mediate the binding of DNA to mica, by forming bridges between the negative charges on the phosphate backbone and the negatively charged mica sites. These bridges are strong, since DNA molecules do not return into the solution after deposition onto the surface under standard conditions (32). Moreover, when deposited on untreated freshly cleaved mica, DNA molecules equilibrate onto the surface as in an ideal 2D solution, while the conformations of DNA molecules deposited on glow-discharged or H$^+$-exchanged mica resemble the projection of the bulk 3D conformations onto a 2D plane (32). The latter results from increasing the interaction energy between DNA and the surface by discharging, soaking, or rinsing the mica plate with de-ionized water, which displaces the metal ions. These stronger binding energies prevent lateral diffusion and equilibration of the DNA molecule.

To our knowledge, the deposition of DNA/H-NS complexes on mica surfaces has not been studied in such detail. It is, however, interesting to note that Dame *et al.* (7) deposited the complexes on freshly cleaved surfaces without prior treatment, representing conditions where naked DNA would equilibrate as in an ideal 2D solution. We therefore investigated the effects of constraining the DNA plasmid to remain close to a plane (see the Methods section and the Supporting Material for more detail on these calculations). Constraining DNA to move in 2D rather than in 3D has a huge impact on the elastic bending energy, the apparent rigidity of the molecule, the number of *cis*- and *trans*-bonds that can be formed, and therefore on the whole compaction dynamics. Fig. 4 shows the time evolution of the number of H-NS dimers bound to DNA obtained from 2D simulations with $G = 2\,k_B T$ (bottom plot) and $G = 4\,k_B T$ (top plot), including the partitioning in dangling dimers and *cis*- and *trans*-binding dimers. Comparison of 2D and 3D simulations indicates that the general trends observed in 3D are preserved in 2D, in particular the swapping of the numbers of *cis*- and *trans*-binding H-NS dimers when switching from $G = 4\,k_B T$ to $G = 2\,k_B T$. However, the number of *trans*-binding dimers is substantially smaller in 2D than in 3D for both values of $G$. For $G = 2\,k_B T$, this number has become so small that it cannot be distinguished anymore in the bottom plot of Fig. 4. When looking at the degree of compaction, measured by the number of DNA sites occupied by *trans*-binding H-NS dimers, it is clear that compaction hardly takes place in 2D for $G = 2\,k_B T$ (bottom plot of Fig. 5), while it is rapid and quite homogeneous for $G = 4\,k_B T$ (top plot of Fig. 5).

The differences between condensed conformations in 2D and 3D are even more striking. For $G = 4\,k_B T$, condensed planar plasmid molecules resemble coil-shaped filaments, caused by the many parallel *trans*-binding H-NS dimers (see Fig. 6), while in 3D they adopt a more globular shape (see the snapshots in Fig. 3 and movie S1). Similarly, for $G = 2\,k_B T$ the (eventually) condensed planar plasmids consist of open loops connected by bridges (see Fig. 6), which differ markedly from the more intricate conformations observed in 3D (see the snapshots in Fig. 3 and movie S2). Interestingly, the condensed conformations computed in 2D for $G = 2\,k_B T$ are quite similar to those observed by Dame *et al.*, who reported that open end-loops are present on all observed DNA/H-NS complexes (7). It is therefore tempting to conclude that the correct value for $G$, *i.e.* the one that correctly predicts the experimental compaction results, is $G = 2\,k_B T$. However, this conclusion is premature because the reproducible observation of open end-loops may merely reflect the higher affinity of the mica surface for complexes with open DNA end-loops than for completely condensed DNA molecules (7). Stated in other words, if coil-shaped filaments detach preferentially from the surface during rinsing, then one would observe only complexes with open end-loops during scanning force microscopy experiments, even if coil-shaped filaments had also formed during the equilibration phase. Simulations with a more realistic description of the interactions between DNA and the surface, thus facilitating a comparison between the surface binding strengths of coiled *versus* open end-looped conformations, would therefore be of great interest to obtain a firm estimate for the value of $G$.

Nevertheless, the results of the 2D simulations agree much better with experimentally observed condensed complexes than those of the 3D simulations. This suggests that relaxation of the DNA/H-NS complexes on a mica surface strongly affects their geometry and that the conformations observed in the course of AFM experiments may consequently not be the mere projections of bulk conformations on the surface, but may instead differ significantly from those that prevail in living cells.

**CONCLUSION**

In this work, we have constructed a coarse-grained model description of the compaction of bacterial DNA by H-NS proteins. The level of coarse-graining of the model is sufficiently high to allow for the numerical integration of trajectories for 10 ms but fine enough to match the experimental observation that each bound H-NS molecule occupies about 15 bp. Results of the simulations highlight the fact that DNA compaction is driven by the subtle equilibrium between several competing factors, including the deformation dynamics of the plasmid and the several binding modes of protein dimers, *i.e.* dangling configuration, *cis-* and *trans*-binding. In particular, the simulations showed that condensed DNA conformations are very sensitive to the binding energy difference between the *cis* and *trans*-binding conformations. If H-NS *cis*-binding dimers are not substantially less stable than *trans*-binding dimers, then *cis*-binding and dangling dimers occupy most of DNA binding sites, and the number of *trans*-binding dimers, equal to the number of bridges between two DNA sites that are at least three beads apart, remains small. This leads to moderately condensed DNA conformations, which are still able to deform strongly. In contrast, if *cis*-binding dimers are substantially less stable than the *trans*-binding ones, then the number of *trans*-binding proteins becomes large, resulting in DNA rapidly acquiring a globular conformation, which changes little in time. Our simulations also pointed out that the conformations of DNA/H-NS complexes are significantly different in 2D and in 3D. This observation suggests that the relaxation of complexes on a mica surface has a strong influence on the observed conformations and, consequently, that conformations observed on mica plates may differ from those that prevail in living cells.

There are several lines along which the model described in this work can be extended and improved.

First of all, a more realistic description of the interactions between the plasmid and the mica surface would be of great interest to confirm that the conformations observed during scanning force microscopy experiments indeed result from the equilibration of globular bulk conformations on the mica surface. To this end, one should first construct a suitable interaction term between DNA and the surface, which would lead to correct predictions for the dynamics of naked DNA deposition. in particular regarding the complete relaxation of DNA at low surface charge density and the freezing of lateral motion at higher charge density. By using this interaction potential for DNA/H-NS complexes, the simulations can shed light on whether H-NS bridges are sufficient to prevent equilibration or whether the bridges that exist prior to deposition break and/or reorganize during deposition and equilibration.

A straightforward extension of this work consists of performing similar simulations close to *in vivo* conditions, instead of the experimental conditions of Dame *et al* (7). As a necessary benchmark to adjust the parameters of the model, the simulations reported in this paper were indeed performed with one 2686 bp plasmid and 224 H-NS dimers to allow for comparison with experiments. In contrast, bacterial DNA may be several millions base pairs long and each cell contains approximately one H-NS dimer per 200 DNA base pairs, instead of one H-NS dimer per 12 DNA base pairs as in Dame *et al* experiment (7). Performing simulations in *in vivo* conditions could provide interesting information, such as for example an estimation of the compaction factor due to H-NS, but will also certainly be very expensive from the computational point of view.

Yet another step closer to simulating the actual bacterial genome would involve including additional different proteins, known to play an important role in the compaction of bacterial DNA, such as HU, IHF, StpA, Dps or FIS. StpA, which is homologous to H-NS, is also able to form bridges between two separate DNA duplexes (34). During the stationary phase, Dps binds to chromosomal DNA, thereby forming a stable complex called a *biocrystal*, but the mode of the DNA/Dps interaction is not yet fully understood (33). Fis induces compaction of DNA essentially due to bending (35). At last, HU and IHF, which are homologous to each other, condense DNA by bending and undertwisting it (36). Introducing all these different proteins into a single model is a formidable task, but one could, as a first step, investigate the essential properties of each of them by using a similar mesoscopic model as used here for H-NS.

## SUPPORTING MATERIAL

An accurate description of the model, 6 figures and 2 movies are available at http://www.biophysj.org/biophysj/supplemental/XXXX

## SUPPORTING CITATIONS

References (37-40) appear in the Supporting Material.

**FIGURE LEGENDS**

**Figure 1** : **Time evolution of the number of H-NS dimers bound to DNA (3D simulations).**

Area charts showing the time evolution of the number of H-NS dimers bound to DNA, and the partitioning into dangling dimers and *cis-* and *trans*-binding dimers, for 3D simulations with $G = 2\,k_{\mathrm{B}}T$ (bottom plot) and $G = 4\,k_{\mathrm{B}}T$ (top plot). The system consists of 224 H-NS dimers and one plasmid with 179 binding sites. Results were averaged over 8 different trajectories starting from thermalized plasmid conformations.

**Figure 2** : **Time evolution of bridged DNA sites (3D simulations).**

Diagrams showing the time evolution of bridged DNA sites for single 3D simulations with $G = 2\,k_{\mathrm{B}}T$ (bottom plot) and $G = 4\,k_{\mathrm{B}}T$ (top plot). DNA sites are numbered from 1 to 179 and reported on the vertical axis. Every 10 μs, the DNA sites occupied by a *trans*-binding H-NS dimer are listed and ticks are placed in the diagram at the corresponding locations.

**Figure 3** : **Time evolution of the radius of gyration of the plasmid molecule (3D simulations).**

The bottom plot shows the time evolution of the radius of gyration of the plasmid molecule for single simulations with $G = 2\,k_{\mathrm{B}}T$ and $G = 4\,k_{\mathrm{B}}T$. The inserts above this plot show the conformations of the plasmid molecule at $t$=5.5 ms, 6.5 ms and 7.5 ms for $G = 2\,k_{\mathrm{B}}T$ (top row) and $G = 4\,k_{\mathrm{B}}T$ (second row). H-NS dimers are not shown for clarity.

**Figure 4** : **Time evolution of the number of H-NS dimers bound to DNA (2D simulations).**

Area charts showing the time evolution of the number of H-NS dimers bound to DNA, and the partitioning into dangling dimers and *cis-* and *trans*-binding dimers, for 2D simulations with $G = 2\,k_{\mathrm{B}}T$ (bottom plot) and $G = 4\,k_{\mathrm{B}}T$ (top plot). The system consists of 224 H-NS dimers and one plasmid with 179 binding sites. Results were averaged over 4 different trajectories starting from circular plasmid conformations.

**Figure 5** : **Time evolution of bridged DNA sites (2D simulations).**

Diagrams showing the time evolution of bridged DNA sites for single 2D simulations with $G = 2\,k_{\mathrm{B}}T$ (bottom plot) and $G = 4\,k_{\mathrm{B}}T$ (top plot). DNA sites are numbered from 1 to 179 and reported on the vertical axis. Every 10 μs, the DNA sites occupied by a *trans*-binding H-NS dimer are listed and ticks are placed in the diagram at the corresponding locations.

**Figure 6** : **Typical 2D plasmid conformations.**

Typical conformations of the plasmid for the trajectories shown in Fig. 5. Snapshots were taken at $t$=5.0 ms (for $G = 2\,k_\mathrm{B}T$ ) and $t$=9.0 ms (for $G = 4\,k_\mathrm{B}T$ ), respectively.

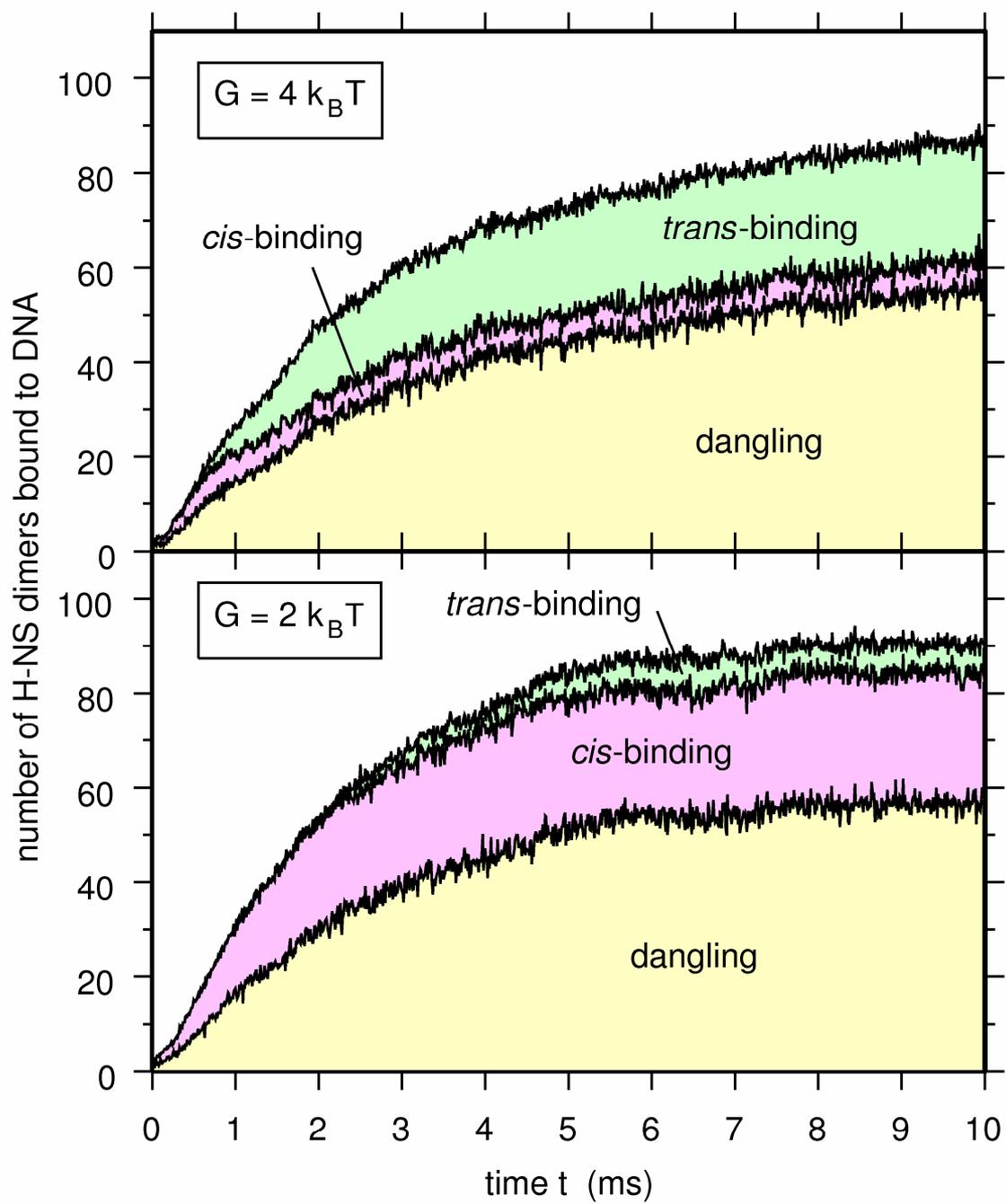

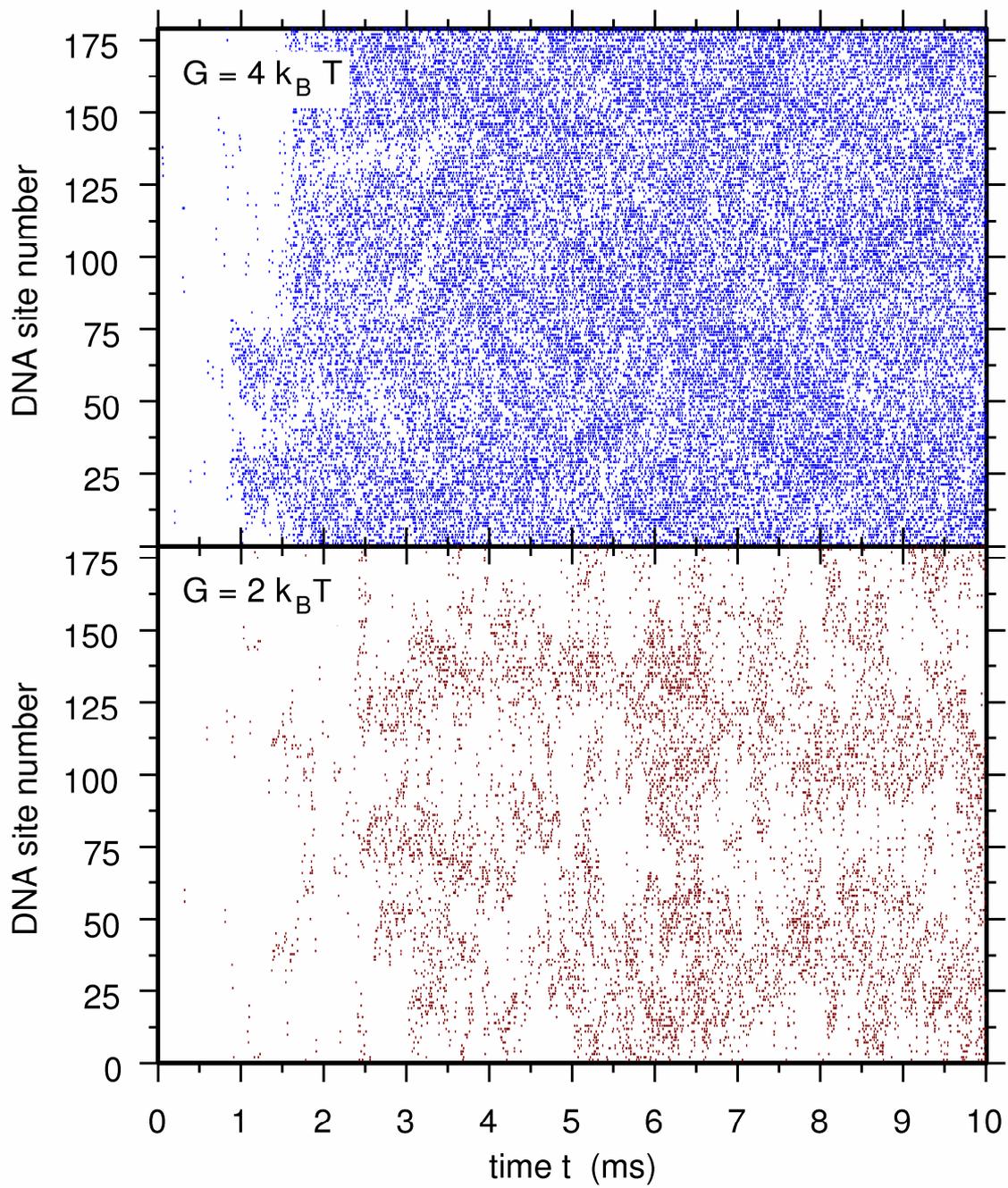

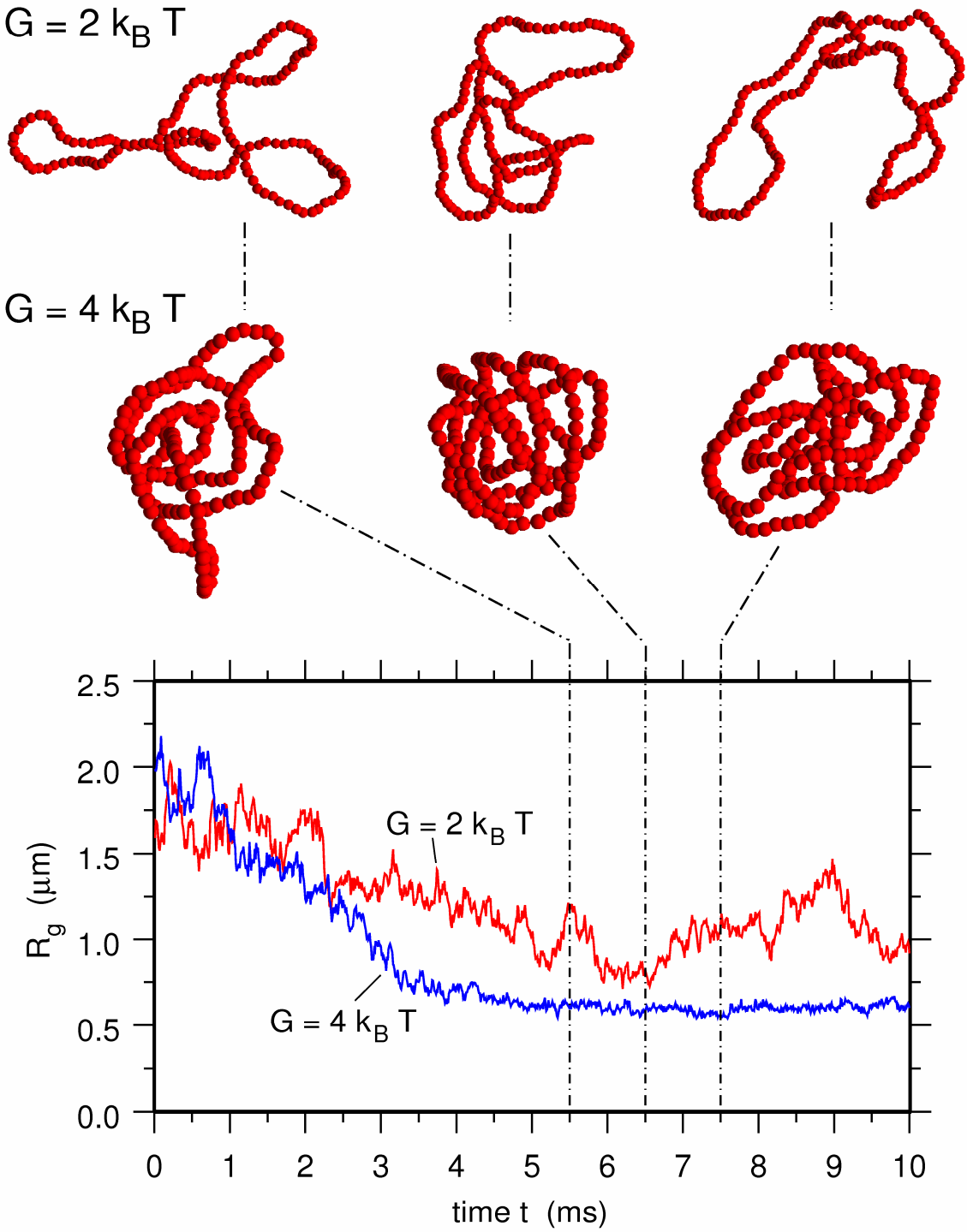

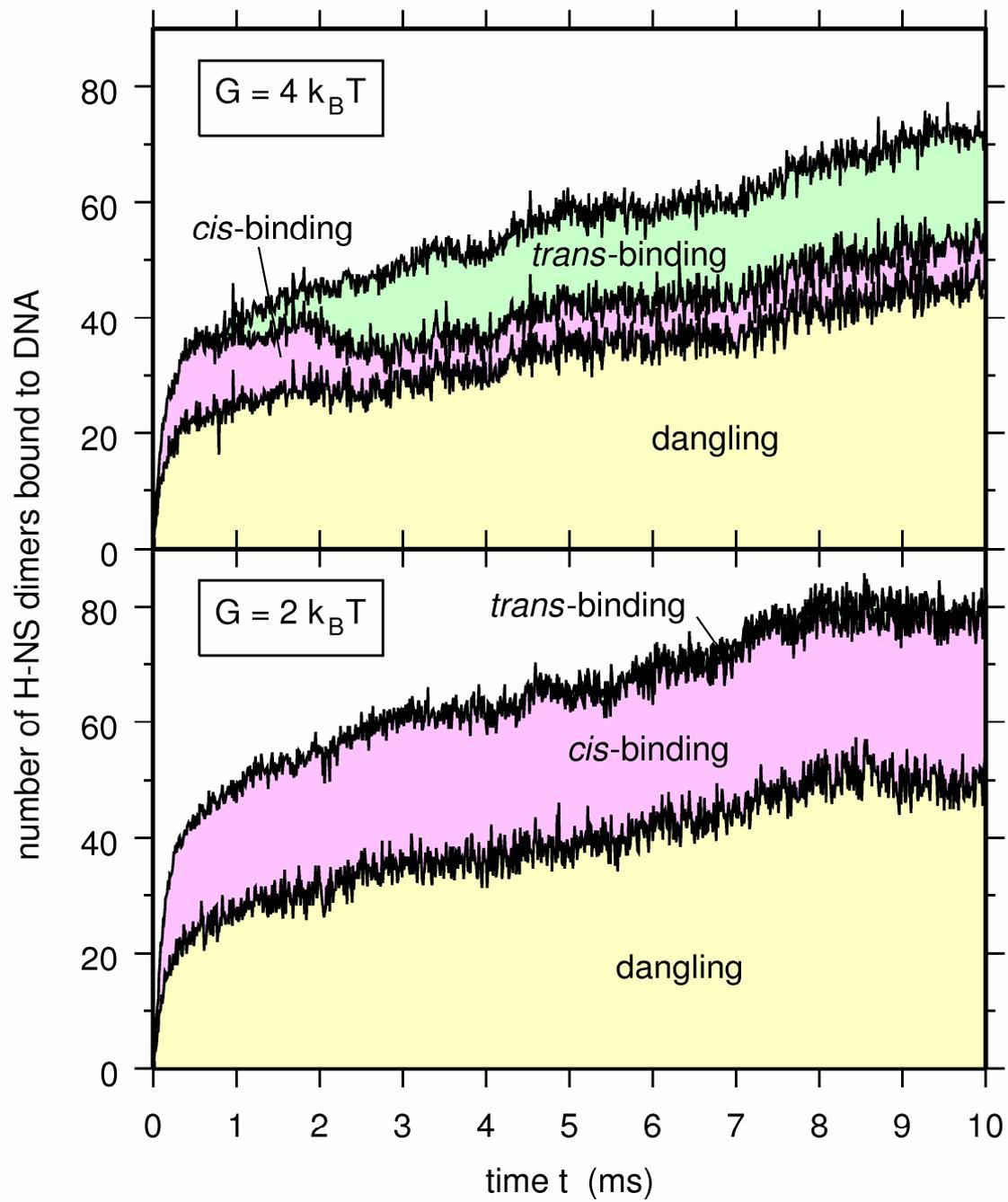

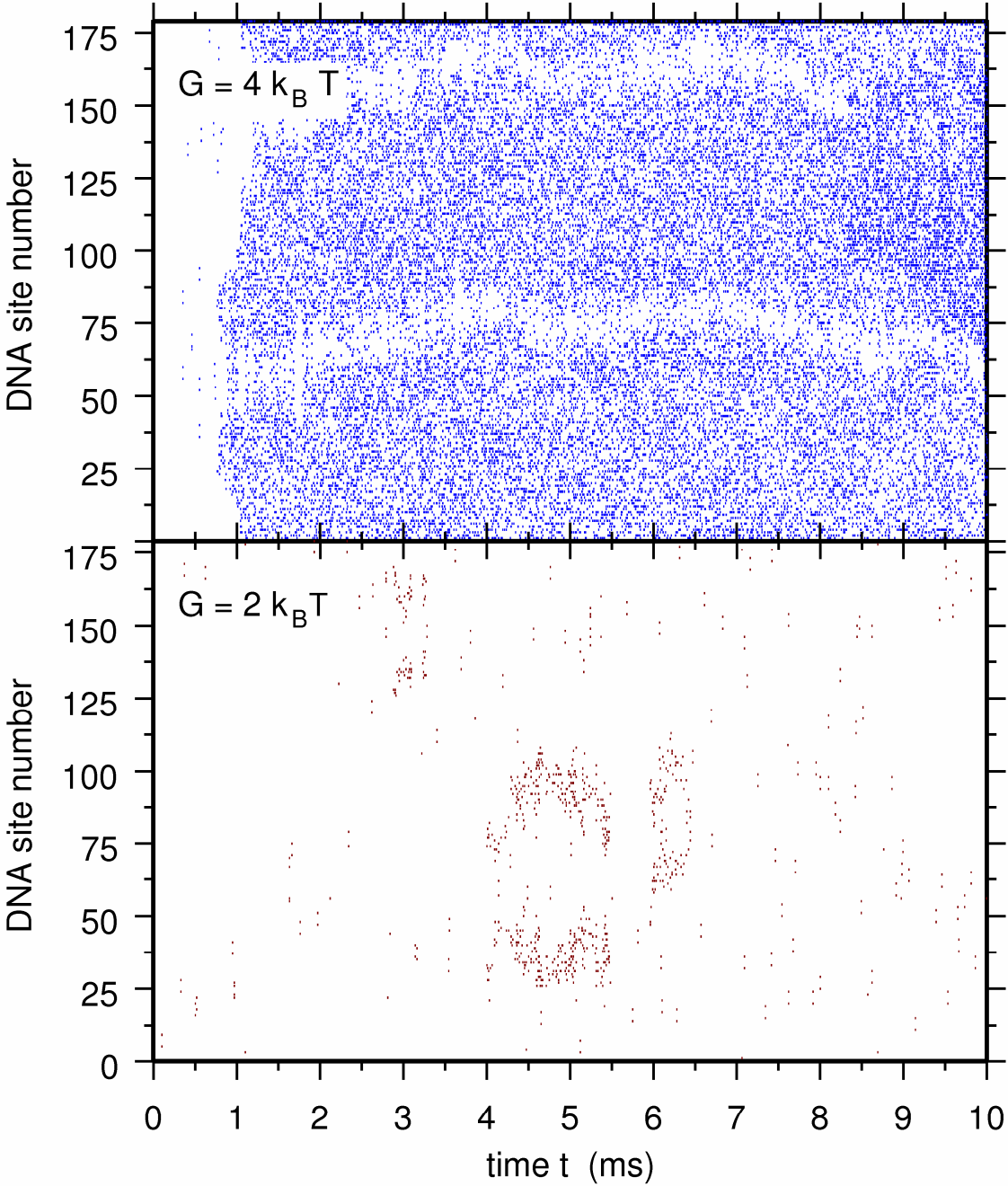

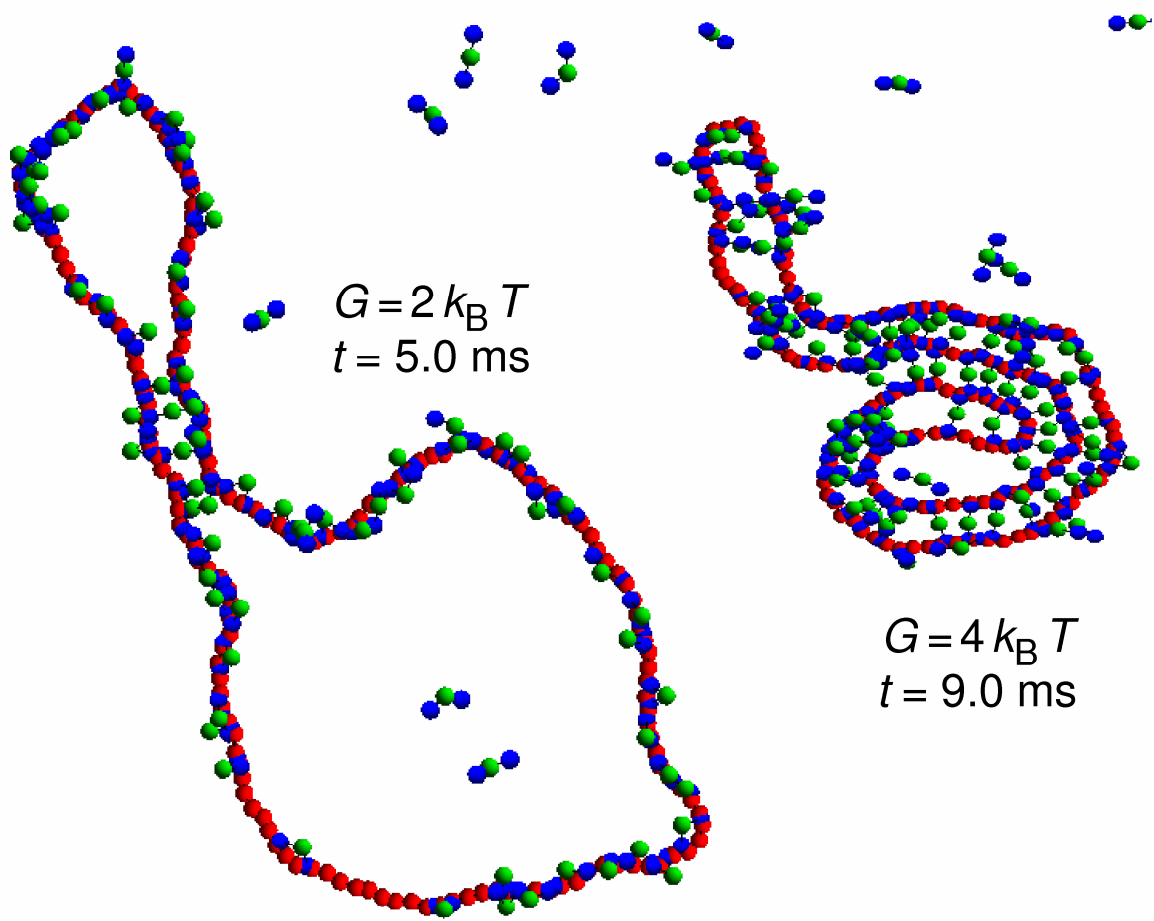

$G = 2\,k_{\mathrm{B}}\,T$
$t = 5.0$ ms

$G = 4\,k_{\mathrm{B}}\,T$
$t = 9.0$ ms

**A model of H-NS mediated compaction of bacterial DNA**
**- Supporting Material -**

Marc JOYEUX
*Laboratoire Interdisciplinaire de Physique (CNRS UMR5588),*
*Université Joseph Fourier Grenoble 1, BP 87, 38402 St Martin d'Hères, France*

and Jocelyne VREEDE
*Van 't Hoff Institute for Molecular Sciences, University of Amsterdam,*
*P.O. Box 94157, 1090 GD Amsterdam, The Netherlands*

## MODEL AND SIMULATIONS

The model was built to mimic the H-NS/DNA incubation conditions used in the work of Dame and co-workers (1). It consists of one pUC19 plasmid and $P$=224 H-NS dimers at a temperature $T$=298 K (1 $k_B T \approx$ 2.48 kJ/mol) enclosed either in a sphere of radius $R_0 = 0.434$ μm or a half-sphere of radius $R_0 = 0.547$ μm. These values of $R_0$ lead to the same plasmid and H-NS concentrations as the 85 ng of DNA with 450 ng of H-NS diluted in 10 μl H-NS BB used in (1). As in previous work (2-5), DNA was modelled as a chain of beads with hydrodynamic radius $a = 1.78$ nm separated at equilibrium by a distance $l_0 = 5.0$ nm. Each bead actually represents 15 base pairs, so that the pUC19 plasmid with 2686 bp was modelled as a cyclic chain of $n$=179 beads. Each bead had an effective charge $\approx -12\,\overline{e}$ placed at its centre, where $\overline{e}$ denotes the absolute charge of the electron. This effective charge corresponds to the product of the known linear charge density of DNA ($-2.43\,\overline{e}$ /nm) by the equilibrium distance $l_0 = 5.0$ nm between two successive beads. Based on crystallographic data (6), each H-NS dimer was modelled as a chain of 3 beads with the same hydrodynamic radius $a = 1.78$ nm separated at equilibrium by a distance $L_0 = 7.0$ nm. An effective charge $-e_{DNA}/3 \approx 4\,\overline{e}$ was placed at the centre of each terminal bead and a charge $2e_{DNA}/3 \approx -8\,\overline{e}$ at the centre of each central bead. The values of these effective charges were estimated by counting the number of positively and negatively charged residues in published crystallographic structures (6).

The potential energy $E_{pot}$ of the system was taken as the sum of 5 terms

$$E_{\text{pot}} = V_{\text{DNA}} + V_{\text{PROT}} + V_{\text{PROT/PROT}} + V_{\text{DNA/PROT}} + V_{\text{wall}} \; , \tag{S1}$$

where $V_{\text{DNA}}$ and $V_{\text{PROT}}$ describe the potential energy of DNA and H-NS dimers, respectively, $V_{\text{PROT/PROT}}$ the interactions between H-NS dimers, $V_{\text{DNA/PROT}}$ the interactions between DNA and H-NS dimers, and $V_{\text{wall}}$ the repulsive wall that maintains H-NS dimers inside the sphere.

$V_{\text{DNA}}$ was expressed, as in previous work (2-5), as a sum of three terms

$$
\begin{aligned}
V_{\text{DNA}} &= E_{\text{s}} + E_{\text{b}} + E_{\text{e}} \\
E_{\text{s}} &= \frac{h}{2} \sum_{k=1}^{n} (l_k - l_0)^2 \\
E_{\text{b}} &= \frac{g}{2} \sum_{k=1}^{n} \theta_k^2 \\
E_{\text{e}} &= e_{\text{DNA}}^2 \sum_{k=1}^{n-2} \sum_{K=k+2}^{n} H(\|\mathbf{r}_k - \mathbf{r}_K\|),
\end{aligned}
\tag{S2}
$$

where $\mathbf{r}_k$ denotes the position of DNA bead $k$, $l_k = \|\mathbf{r}_k - \mathbf{r}_{k+1}\|$ the distance between two successive beads, $\theta_k = \arccos((\mathbf{r}_k - \mathbf{r}_{k+1})(\mathbf{r}_{k+1} - \mathbf{r}_{k+2})/(\|\mathbf{r}_k - \mathbf{r}_{k+1}\| \|\mathbf{r}_{k+1} - \mathbf{r}_{k+2}\|))$ the angle formed by three successive beads, and $H$ is the function defined according to

$$H(r) = \frac{1}{4\pi\varepsilon r} \exp\left(-\frac{r}{r_D}\right). \tag{S3}$$

$E_{\text{s}}$ is the bond stretching energy. In fact, this is a computational device without any biological meaning aimed at avoiding a rigid rod description. The stretching force constant was fixed at $h = 100 \, k_{\text{B}}T / l_0^2$ (see the discussion in Ref. (2) for this choice for $h$). $E_{\text{b}}$ is the elastic bending potential. The bending rigidity constant, $g = 9.82 \, k_{\text{B}}T$, was chosen so as to provide the correct persistence length for DNA, which is 50 nm, equivalent to 10 beads (2,7). $E_{\text{e}}$ is a Debye-Hückel potential, which describes repulsive electrostatic interactions between DNA beads (2,8,9). In Eq. S3, $r_D = 3.07$ nm stands for the Debye length at 0.01 M salt concentration of monovalent ions (2) and $\varepsilon = 80 \, \varepsilon_0$ for the dielectric constant of the buffer. Note that electrostatic interactions between nearest-neighbours are not included in the expression of $E_{\text{e}}$ in Eq. S2, because these nearest-neighbour interactions are accounted for in the stretching and bending terms.

$V_{\text{PROT}}$ was similarly taken as the sum of stretching and bending contributions

$$V_{\text{PROT}} = E_s^{(P)} + E_b^{(P)}$$

$$E_s^{(P)} = \frac{h}{2} \sum_{j=1}^{P} (L_{j,1} - L_0)^2 + (L_{j,2} - L_0)^2 \tag{S4}$$

$$E_b^{(P)} = \frac{G}{2} \sum_{j=1}^{P} \Theta_j^2,$$

where $L_{j,1}$, $L_{j,2}$, and $\Theta_j$, denote the distances between the terminal beads and the central bead and the angle formed by the three beads for the $j^{\text{th}}$ H-NS dimer. The value of the bending rigidity $G$ was estimated by noticing that the N-terminal oligomerization domain and the C-terminal DNA-binding domain of H-NS are linked principally by a long α-helix. Since the measured persistence length $\xi$ of α-helices is close to 15 nm (10) and is related to the bending rigidity through $G = k_B T \xi / L_0$, it was estimated that $G$ is of the order of a few $k_B T$ for H-NS. Application of the formula leads to $G \approx 2 k_B T$, but this value can of course only be considered as a rough estimate. As discussed in more detail below and in the main text, we actually used two different values for $G$ in the simulations, namely $G = 2\, k_B T$ and $G = 4\, k_B T$, because these values lead to different amounts of H-NS binding in *trans* and, as a consequence, to different levels of DNA compaction.

The interaction between H-NS dimers, $V_{\text{PROT/PROT}}$, was taken as the sum of (attractive or repulsive) electrostatic terms and (repulsive) excluded volume terms, with the latter ones only contributing if the corresponding electrostatic interactions are attractive, *i.e.* between the terminal beads $m=1$ and $m=3$ of one dimer and the central bead $m=2$ of the other dimer

$$V_{\text{PROT/PROT}} = E_e^{(P/P)} + E_{ev}^{(P/P)}$$

$$E_e^{(P/P)} = \sum_{j=1}^{P} \sum_{m=1}^{3} \sum_{J=j+1}^{P} \sum_{M=1}^{3} e_{jm} e_{JM} H \left( \left\| \mathbf{R}_{jm} - \mathbf{R}_{JM} \right\| \right) \tag{S5}$$

$$E_{ev}^{(P/P)} = \chi \sum_{j=1}^{P} \sum_{\substack{J=1 \\ J \neq j}}^{P} \left( \left| \frac{e_{j1} e_{J2}}{e_{\text{DNA}}^2} \right| F \left( \left\| \mathbf{R}_{j1} - \mathbf{R}_{J2} \right\| \right) + \left| \frac{e_{j3} e_{J2}}{e_{\text{DNA}}^2} \right| F \left( \left\| \mathbf{R}_{j3} - \mathbf{R}_{J2} \right\| \right) \right),$$

where $\mathbf{R}_{jm}$ denotes the position of bead $m$ of protein dimer $j$, $e_{jm}$ the charge placed at its centre, $\chi$ is a constant equal to $\chi = 0.15\, k_B T$, and $F$ is the function defined according to if $r \leq 2^{3/2} a$ : $F(r) = 4 \left( \left( \frac{2a}{r} \right)^4 - \left( \frac{2a}{r} \right)^2 \right) + 1$

if $r > 2^{3/2} a$ : $F(r) = 0$ . $\tag{S6}$

It should be noted that electrostatic attractive interactions between the terminal beads of one H-NS dimer and the central bead of another H-NS dimer are too weak to allow for the formation of long-lived bonds between these two dimers. Consequently, the model proposed here does not take oligomerization of the dimers into account. As mentioned in the Introduction section in the main text, H-NS is believed to be functional as a dimer, but trimers, tetramers and larger oligomers have been observed in solution under different conditions. If future work proves that these higher-order oligomers play an important role in H-NS mediated compaction of bacterial DNA, then the $V_{\text{PROT/PROT}}$ interaction term will have to be adapted to take oligomerization into account.

The interaction between DNA and protein dimers, $V_{\text{DNA/PROT}}$, was similarly taken as the sum of (attractive or repulsive) electrostatic terms and (repulsive) excluded volume terms, with the latter ones only contributing if the corresponding electrostatic interactions are attractive, *i.e.* between DNA beads and terminal protein beads $m=1$ and $m=3$,

$$V_{\text{DNA/PROT}} = E_e^{(\text{DNA/P})} + E_{ev}^{(\text{DNA/P})}$$

$$E_e^{(\text{DNA/P})} = \sum_{j=1}^{P} \sum_{m=1}^{3} \sum_{k=1}^{n} e_{jm} e_{\text{DNA}} H(\|\mathbf{R}_{jm} - \mathbf{r}_k\|) \qquad (S7)$$

$$E_{ev}^{(\text{DNA/P})} = \chi \sum_{j=1}^{P} \sum_{k=1}^{n} \left( \left| \frac{e_{j1}}{e_{\text{DNA}}} \right| F(\|\mathbf{R}_{j1} - \mathbf{r}_k\|) + \left| \frac{e_{j3}}{e_{\text{DNA}}} \right| F(\|\mathbf{R}_{j3} - \mathbf{r}_k\|) \right).$$

The confining sphere is large enough to fit the plasmid, whatever its current geometry. Indeed, the diameter of the sphere is as large as 0.868 μm for 3D simulations and 1.094 μm for 2D simulations, while the maximum length reached by the completely stretched plasmid is 0.450 μm. Since the centre of the sphere was adjusted at each time step to coincide with the centre of mass of the DNA molecule (see below), DNA beads could not exit the sphere. In order to restrain protein beads from exiting the sphere, we introduced a repulsive wall $V_{\text{wall}}$, which acts on the protein beads that move outside the radius of the sphere, $R_0$, and repel them back into the sphere. $V_{\text{wall}}$ was taken as a sum of repulsive terms

$$V_{\text{wall}} = 10 \, k_B T \sum_{j=1}^{P} \sum_{m=1}^{3} f(\|\mathbf{R}_{jm}\|) , \qquad (S8)$$

where $f$ is the function defined according to

if $r \leq R_0$ : $f(r) = 0$

if $r > R_0$ : $f(r) = \left( \frac{r}{R_0} \right)^6 - 1$ . $\qquad (S9)$

At this point, it is important to emphasize that, within this model, interactions between DNA and H-NS dimers are essentially driven by the constant $\chi$ in Eq. S7. The value $\chi = 0.15\,k_B T$ was chosen because it leads to a change in enthalpy $\Delta H$ of 11.1 $k_B T$ on forming a complex between DNA and an H-NS monomer, which is comparable to experimentally determined values (11) (see Fig. S1). Moreover, the energy landscape for *cis*-binding of H-NS can be visualized by setting one H-NS terminal bead at the minimum energy location and varying the position of the other terminal bead while keeping the bond lengths at their equilibrium value $L_0$. The result is shown in Fig. S2 for $G = 2\,k_B T$ and in Fig. S3 for $G = 4\,k_B T$. The maximum *cis*-binding enthalpy change is 19.2 $k_B T$ for $G = 2\,k_B T$ and 16.8 $k_B T$ for $G = 4\,k_B T$.

The dynamics of the system was investigated by integrating numerically the Langevin equations of motion with kinetic energy terms neglected. Practically, the updated position vector for each bead (whether DNA or protein), $\mathbf{r}^{(n+1)}$, was computed from the current position vector, $\mathbf{r}^{(n)}$, according to

$$\mathbf{r}^{(n+1)} = \mathbf{r}^{(n)} + \frac{\Delta t}{6\pi\eta\,a}\mathbf{F}^{(n)} + \sqrt{\frac{2\,k_B T\,\Delta t}{6\pi\eta\,a}}\,\xi^{(n)}, \qquad (S10)$$

where $\Delta t = 20$ ps is the integration time step, $\mathbf{F}^{(n)}$ the collective vector of inter-particle forces arising from the potential energy $E_{pot}$, $\xi^{(n)}$ a vector of random numbers extracted at each step $n$ from a Gaussian distribution of mean 0 and variance 1, and $\eta = 0.00089$ Pa s, the viscosity of the buffer at 298 K. After each integration step, the position of the centre of the sphere was adjusted so as to coincide with the centre of mass of the DNA molecule. Two different sets of simulations were performed. The first set of simulations, labelled 3D, aimed at mimicking the motion of the system in bulk buffer, by allowing all molecules to move freely in the sphere. The second set of simulations, labelled 2D, aimed at mimicking DNA molecules deposited on cleaved mica surfaces. To this end, the DNA molecule was constrained to remain in the neighbourhood of the $z=0$ surface, in the $z$-range extending from 0 to 0.5 nm, while the $z$-coordinate of each H-NS bead was constrained to remain positive. All simulations were run for 10 ms.

Analysis of the simulations consisted in tracking every 10 µs the number of bound H-NS dimers. It was considered that an H-NS dimer was bound to DNA if it had at least one bead within a distance of 2 nm from a DNA bead. The nature of the bound H-NS dimers was also tracked. A dimer was considered to be bound in *cis* if the two terminal beads were interacting with DNA beads separated by at most two DNA beads. If instead, the H-NS dimer was interacting with two DNA beads separated by (strictly) more than two DNA beads, the H-

NS/DNA interaction was counted as a *trans* bond. Finally, the dimer was considered to be dangling, if only one bead of the dimer was interacting with the DNA. As illustrated in Fig. S4, there is actually not much freedom in the choice of the distance threshold. Indeed, for thresholds smaller than 2.0 nm, many bound proteins are missed because of the steep repulsive wall that surrounds DNA beads (see Figs. S1-S3). On the other hand, the threshold cannot be larger than half the separation between two DNA beads, that is, 2.5 nm. The 2 nm threshold was chosen, because Fig. S4 suggests that a certain number of dangling dimers are incorrectly reassigned as *cis*-binding dimers when the threshold is increased from 2.0 to 2.5 nm. Most importantly, the estimation for the number of *trans*-binding dimers remains essentially constant in this latter range of threshold values.

The radius of gyration of the plasmid molecule, *i.e.* the square root of the trace of its gyration tensor, was also computed every 10 μs.

The results of the simulations are discussed and illustrated in the main text. Fig. S5 additionally shows the time evolution of the occupancy ratio of DNA sites for $G = 2\,k_\mathrm{B}T$ and $G = 4\,k_\mathrm{B}T$ and Fig. S6 the time evolution of the radius of gyration of the plasmid for eight simulations with $G = 2\,k_\mathrm{B}T$ and $G = 4\,k_\mathrm{B}T$.

**Movie S1** (.avi file, 4862 Ko) : **The condensed plasmid for $G$=4 $k_BT$.**

This movie shows the evolution of the plasmid conformations in the time interval 5.5-7.5 ms for the 3D simulation with $G = 4\,k_BT$ , also shown in Fig. 2 (top). The plasmid chain is shown as red beads and the H-NS dimers as green (central) and blue (terminal) beads. Frames were generated every 10 µs.

**Movie S2** (.avi file, 6129 Ko) : **The condensed plasmid for $G$=2 $k_BT$.**

This movie shows the evolution of the plasmid conformations in the time interval 5.5-7.5 ms for the 3D simulation with $G = 2\,k_BT$ also shown in Fig. 2 (bottom). The plasmid chain is shown as red beads and the H-NS dimers as green (central) and blue (terminal) beads. Frames were generated every 10 µs.

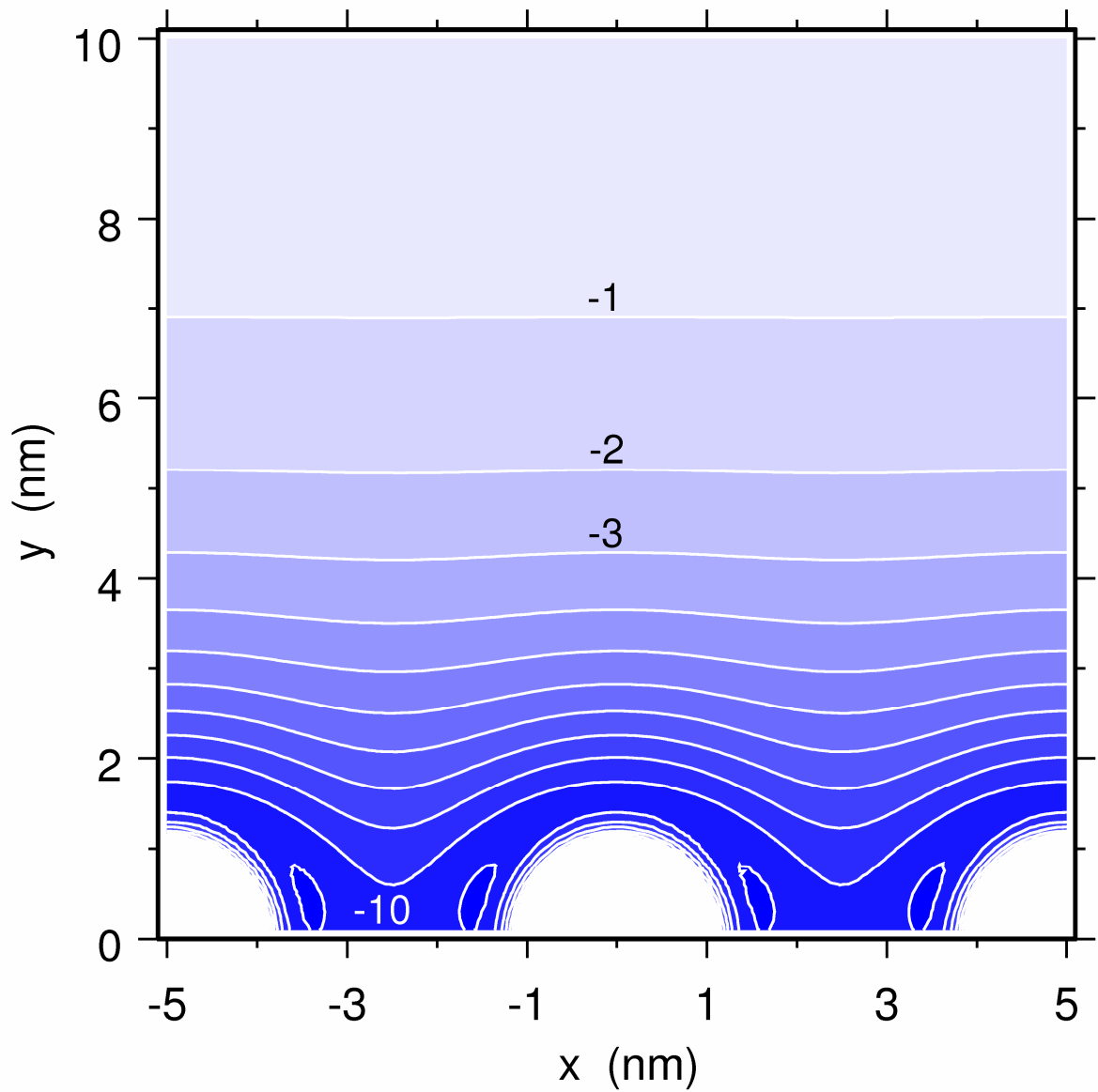

**Figure S1 :** Potential energy $V_{\text{DNA/PROT}}$ seen by a linear H-NS dimer oriented parallel to the $y$ axis and with equilibrium bond lengths $L_0$. $(x,y)$ denote the coordinates of the H-NS terminal bead closest to DNA. DNA beads are located on the $x$ axis at positions ..., -5, 0, 5, ... nm. Contours start at -1 $k_{\text{B}}T$ and are separated by 1 $k_{\text{B}}T$, that is, approximately 2.48 kJ/mol at 298 K.

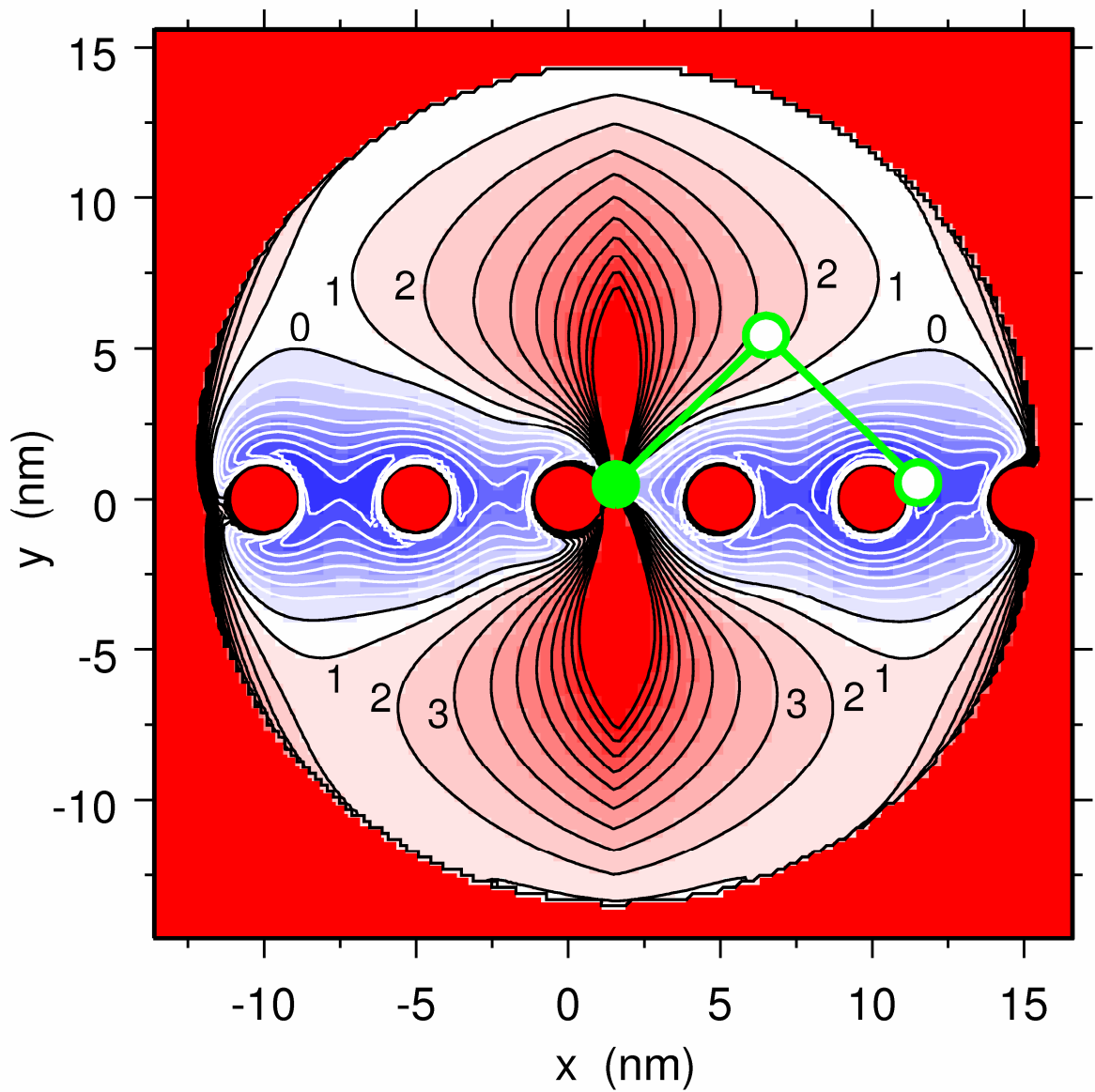

**Figure S2 :** Potential energy $V_{\text{DNA/PROT}}$ seen by an H-NS dimer with equilibrium bond lengths $L_0$ and terminal bead $m=1$ (shown as a filled green circle) sitting on the minimum energy position, for protein bending rigidity $G=2\,k_{\text{B}}T$. $(x,y)$ denotes the coordinates of H-NS terminal bead $m=3$. DNA beads are located on the $x$ axis at positions ...-10, - 5, 0, 5, 10, 15... nm, as in Fig. S1. Energy values are plotted relative to the monomer minimum at -11.1 $k_{\text{B}}T$ (Fig. S1). Blue regions, with negative relative energies, therefore correspond to dimers that are more stable than the equilibrium monomer, and red regions to less stable dimers. Contours are separated by 1 $k_{\text{B}}T$, that is, approximately 2.48 kJ/mol at 298 K. An H-NS dimer with minimum *cis*-binding energy conformation is shown in green.

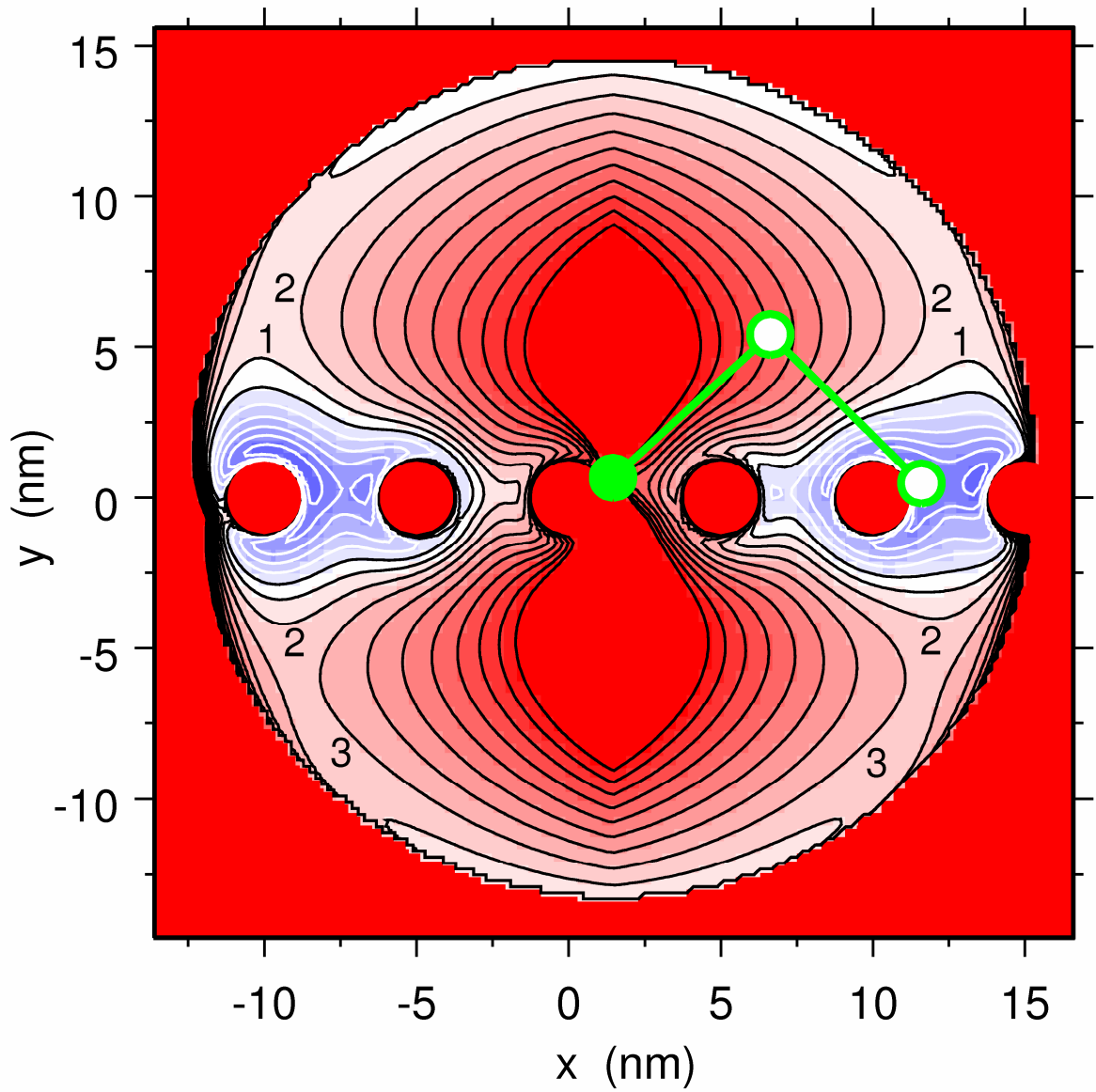

**Figure S3 :** Same as Fig. S2 but for protein bending rigidity $G = 4\,k_{\mathrm{B}}T$ .

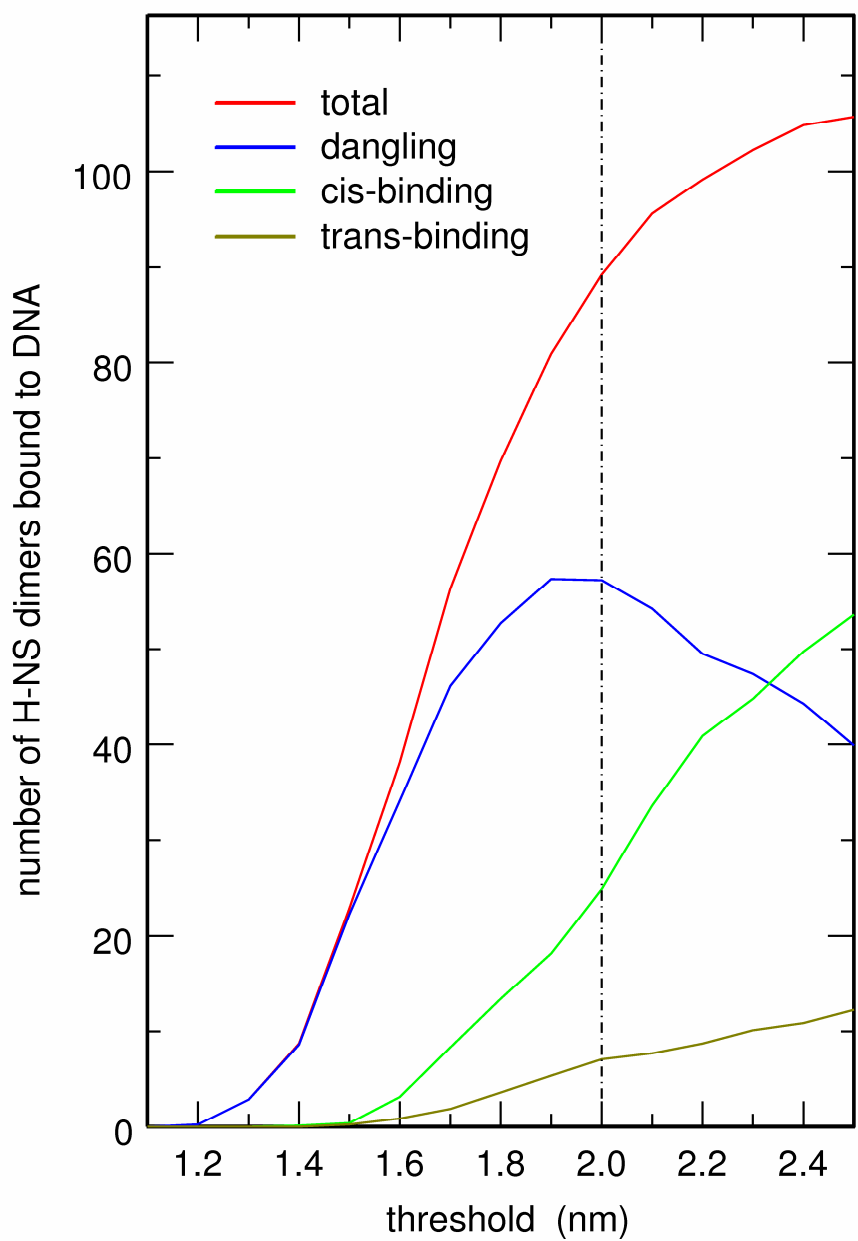

**Figure S4 :** Plot of the average number of H-NS dimers bound to DNA (red line), as well as the partition into dangling dimers (blue line), *cis*-binding dimers (green line), and *trans*-binding dimers (brown line), as a function of the threshold distance used to decide whether two beads are interacting together or not. Results presented in this paper were actually obtained with a 2.0 nm threshold (vertical dot-dashed line). This plot was computed from the conformations at final time (10 ms) for the eight 3D simulations with $G = 2\,k_{\mathrm{B}}T$ .

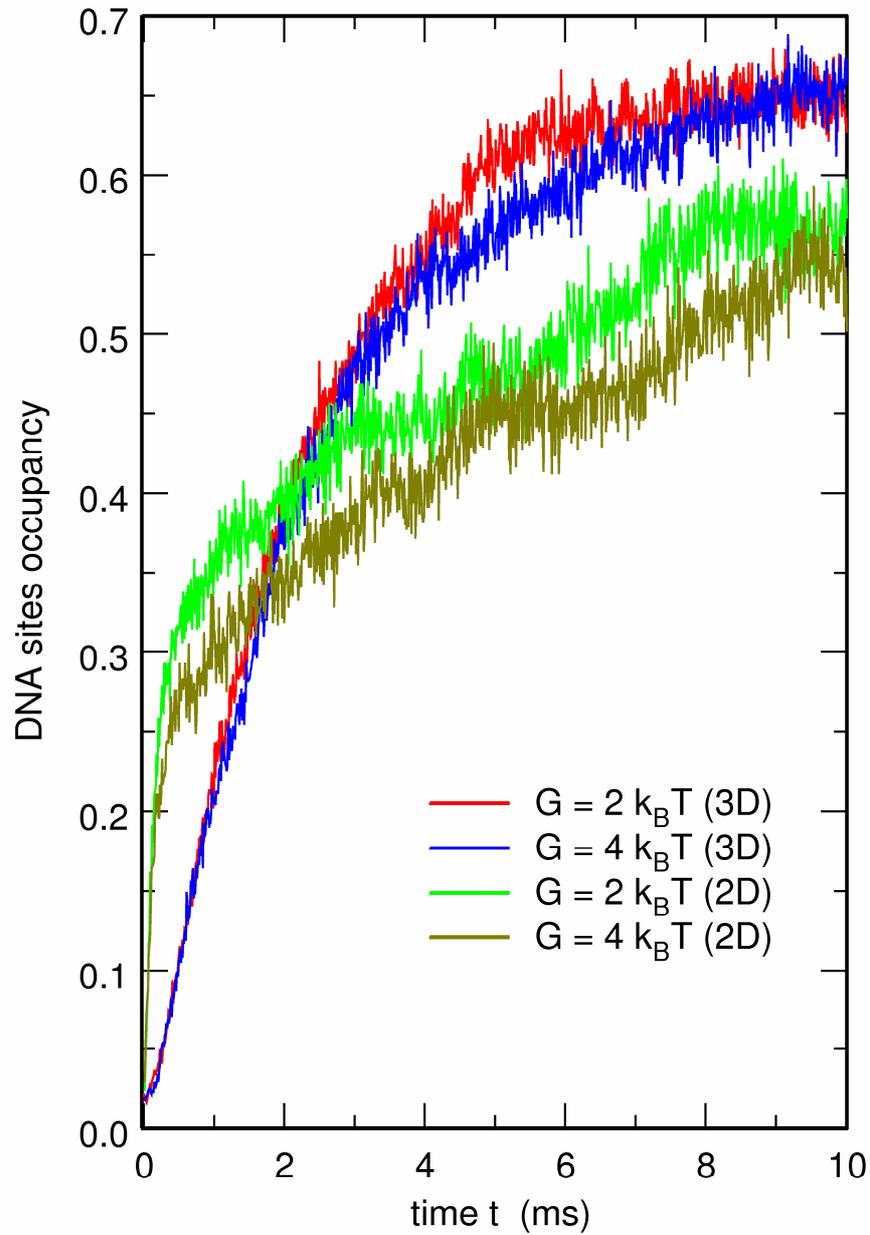

**Figure S5 :** Time evolution of the ratio of DNA sites bound to an H-NS dimer to the total number of DNA sites, for $G = 2\,k_{\mathrm{B}}T$ and $G = 4\,k_{\mathrm{B}}T$, and for 2D as well as 3D simulations. Note that the plasmid model contains 179 sites to which H-NS dimers can bind. 2D (respectively, 3D) results were averaged over 4 (respectively, 8) different trajectories.

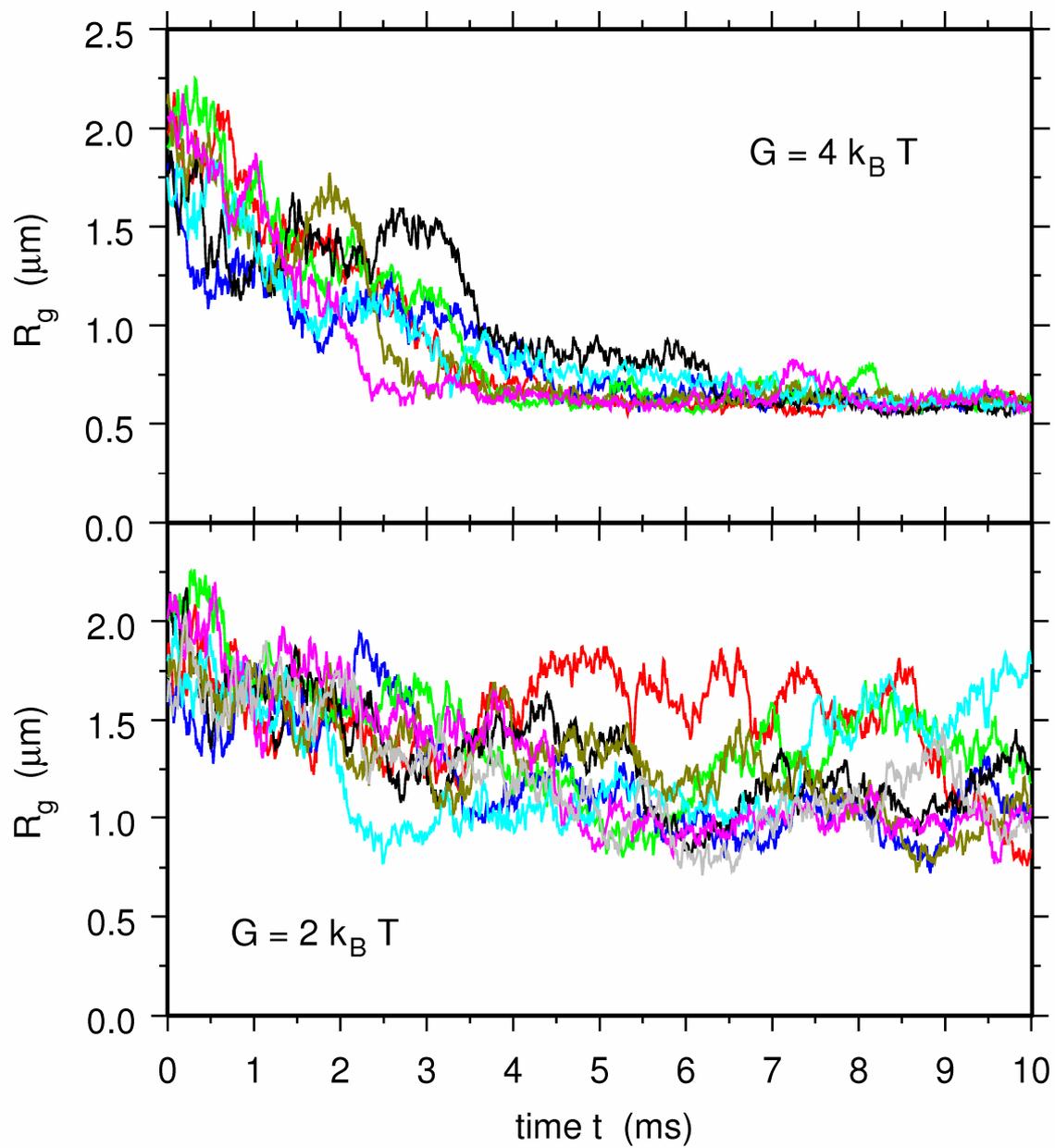

**Figure S6** : The bottom and top plots show the time evolution of the radius of gyration of the plasmid molecule for eight simulations with $G = 2\,k_B T$ and $G = 4\,k_B T$, respectively.